# DIRAC – Distributed Infrastructure with Remote Agent Control

A.Tsaregorodtsev, V.Garonne
*CPPM-IN2P3-CNRS, Marseille, France*

J. Closier, M. Frank, C. Gaspar, E. van Herwijnen, F. Loverre, S. Ponce
*CERN, Geneva, Switzerland*

R.Graciani Diaz
*Univ. Barcelona, Spain*

D. Galli, U. Marconi, V. Vagnoni
*INFN, Bologna, Italy*

N. Brook
*Wills Physics Laboratory, Bristol, UK*

A. Buckley, K. Harrison
*Univ. Cambridge, UK*

M.Schmelling
*GRIDKA, Karlsruhe, Germany*

U.Egede
*Imperial College, Univ. London, UK*

A. Bogdanchikov
*INP, Novosibirsk, Russia*

I. Korolko
*ITEP, Moscow, Russia*

A. Washbrook, J.P.Palacios
*Univ. Liverpool, UK*

S. Klous
*Nikhef and Vrije Universiteit, Amsterdam, Netherlands*

J.J.Saborido
*Univ. Santiago de Compostela, Spain*

A. Khan
*ScotGrid, Edinburgh, UK*

A.Pickford
*ScotGrid, Glasgow, UK*

A. Soroko
*Univ. Oxford, UK*

V. Romanovski
*IHEP, Protvino, Russia*

G.N. Patrick, G.Kuznetsov
*RAL, Oxon, UK*

M. Gandelman
*UFRJ, Rio de Janeiro, Brazil*

Presented by A. Tsaregorodtsev, atsareg@in2p3.fr

This paper describes DIRAC, the LHCb Monte Carlo production system. DIRAC has a client/server architecture based on: Compute elements distributed among the collaborating institutes; Databases for production management, bookkeeping (the metadata catalogue) and software configuration; Monitoring and cataloguing services for updating and accessing the databases. Locally installed software agents implemented in Python monitor the local batch queue, interrogate the production database for any outstanding production requests using the XML-RPC protocol and initiate the job submission. The agent checks and, if necessary, installs any required software automatically. After the job has processed the events, the agent transfers the output data and updates the metadata catalogue. DIRAC has been successfully installed at 18 collaborating institutes, including the DataGRID, and has been used in recent Physics Data Challenges. In the near to medium term future we must use a mixed environment with different types of grid middleware or no middleware. We describe how this flexibility has been achieved and how ubiquitously available grid middleware would improve DIRAC.

**TUAT006**

# 1 INTRODUCTION

LHCb is one of the four future experiments in High Energy Physics at the Large Hadron Collider (LHC) at CERN, Geneva [1]. It will attempt to answer the most challenging questions on the nature of fundamental particles, the origin of the CP violation phenomenon underlying the asymmetry between matter and antimatter. To achieve its goals the LHCb experiment will record an unprecedented amount of the data going up to 2 PB's per year. The data will be distributed in many laboratories in Europe and will be analyzed by the international collaboration of scientists. Therefore, there is a clear need for a distributed hierarchical computing system which will allow to effectively share the data and resources necessary to process them. Many projects are currently in progress to build distributed computing grids. Among them, the EU funded EDG (European Data Grid) project was initiated to provide a solution for such data intensive distributed computing infrastructure.

However, the design of the LHCb experiment requires a large amount of simulation of the physics processes inside the detector. This modeling already now requires a large number of computing resources to work together in a coherent production system. Therefore, the LHCb experiment has built its own distributed production system to satisfy its needs in generation of a large volume of Monte-Carlo simulation data. The system is called DIRAC that stands for the "Distributed Infrastructure with Remote Agent Control".

The DIRAC distributed production system has the following functionality:
- Definition of production tasks
- Software installation on production sites
- Job scheduling and monitoring
- Data bookkeeping and replica management

Many production operations are automated in order to minimize the interventions by local production managers to maintain the system. This is an important requirement for the LHCb considering limited dedicated manpower.

DIRAC has been successfully installed at 18 collaborating institutes and has been also interfaced with the DataGRID project testbed [2]. In the latter case the DataGRID was used as one single production site.

DIRAC was used for the LHCb Physics Data Challenge recently with very good results. Its architecture allowed to fully exploit the available distributed computing resources. The success of the DIRAC system is largely due to the job scheduling mechanism that can be characterized as a "pull" approach.

In the following Section 2 describes the architecture of the system and its main components. Sections 3 gives details on production operations and experience gained with the system. Section 4 describes how DataGrid was integrated into DIRAC, and the tests that were done to study the DataGrid as a production environment. In Section 5 we summarize some of the lessons learnt that will be necessary to take into account in the further development.

# 2 ARCHITECTURE

The design of the production system starts with the choice of necessary components that should cooperate to achieve the production goals. Among these components, the job scheduling (or workload management) system is crucial for the overall performance. Several approaches are possible for the implementation of this component.

## 2.1 "Pull" versus "Push" job scheduling paradigm

One of the choices that should be made in the design of a job scheduling component is whether the scheduler is realizing the "push" or "pull" paradigm.

In the "push" paradigm the scheduler is using the information about the availability and status of the computing resources in order to find the best match to a particular job requirements. Then the job is sent to the chosen computing element for execution. So, in this case the scheduler is an active component whereas the computing element is passive.

In the "pull" paradigm, it is the computing resource that is actively seeking tasks to be executed. The jobs are first accumulated by a production service, validated and put into a waiting queue. Once a computing resource is available, it sends a request for a work to be done to the production service. The production service chooses a job according to the resource capabilities and then serves it in response to the request.

There are advantages and disadvantages in both approaches. In the "push" approach the information about the dynamic status of all the resources is usually collected in one place and for each job a best possible choice can be done. This is fine. But the number of matching operations to be done is proportional to the product of the numbers of resources and jobs. This leads to scalability problems since both numbers will be constantly increasing.

In the "pull" approach the matching is done on demand of a computing resource and for this resource only. This is certainly a more scalable solution although in each scheduling act the choice of a job-resource pair might be suboptimal. In addition there are other advantages as well:
- It is easier to achieve efficient usage of the available power because an idle resource manifests itself immediately.
- The load balancing is also achieved naturally since the more powerful resource will simply request jobs more frequently.
- It is easier to incorporate new production sites since little or no information about them is needed at the central production service.

In developing the DIRAC MC production system we have chosen the "pull" approach.

**TUAT006**

## 2.2 DIRAC architecture components

The DIRAC architecture is presented in Figure 1. It consists of central services: Production, Monitoring and Bookkeeping services, and of software processes called Production Agents that are permanently running on each production site. The central services have facilities for preparing productions, monitoring the jobs execution and bookkeeping of the jobs parameters. The Agents are examining the status of the local production queues. If the local resources are capable of accepting the LHCb workload the Agent is sending a request for outstanding jobs to the central Production service and ensures the execution and monitoring of the job received. After the job has processed the events, the Agent transfers the output data and updates the metadata catalog.

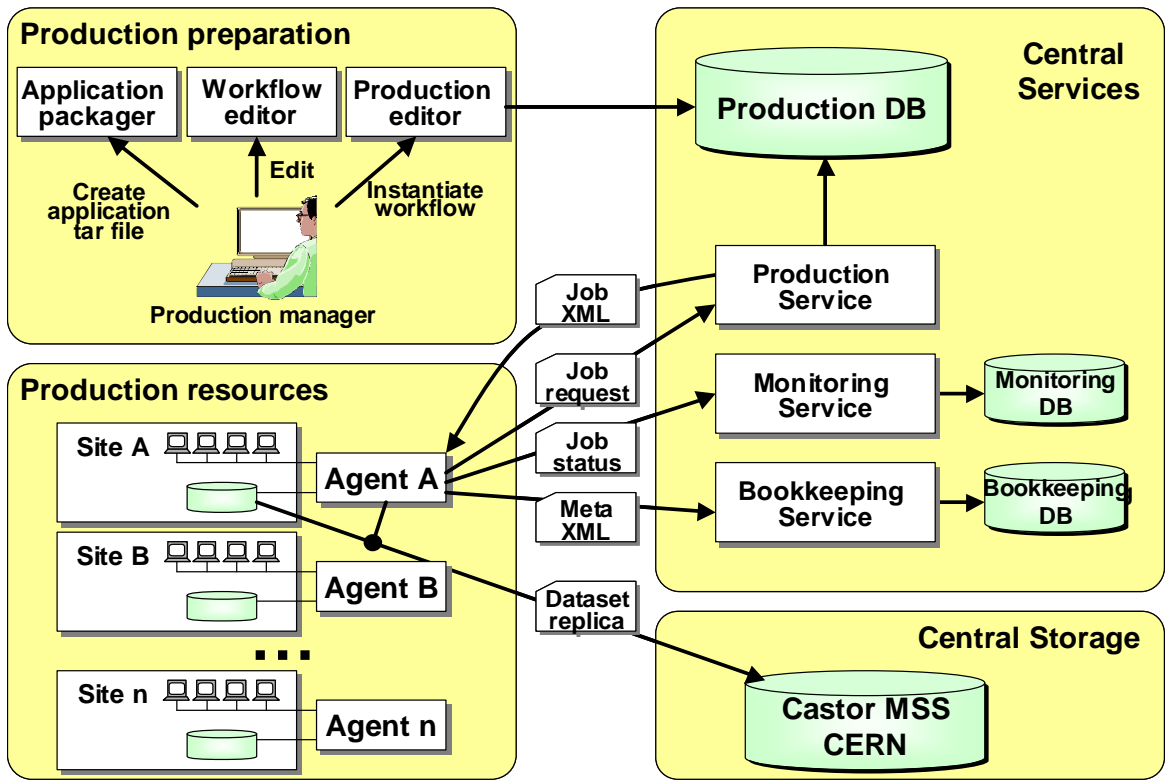

Figure 1: The DIRAC architecture

## 2.3 Software packaging

Before the production can start, the production application software should be prepared for shipping and installation on a production site. It was an important requirement for the DIRAC system to be able to install new versions of the production software immediately after the release is done by an LHCb software manager. Therefore, a utility was provided to prepare a binary package for each application that contains all the necessary libraries to run it. With such packaging we do not make any assumptions of any software preinstalled at a production site. The resulting binary tar file contains a setup script also generated automatically to bootstrap the application on a working node before it runs. The LHCb applications are based on the GAUDI framework [3] and the software is maintained with the help of the CMT code management tool [4]. This allows to fully automate the application packaging.

## 2.4 Production database

Production database is the core component of the central services. All the information describing the production tasks as well as job status parameters are stored in the Production database. It is implemented as an Oracle database running on a CERN server.

## 2.5 Production preparation phase

While the production preparation phase all the necessary software and production tasks are made readily available to the requests of Production Agents. Ideally, this is the only time where the central production manager intervenes. Once the production tasks are defined, their execution is completely automated.

### 2.5.1 Workflow definition

The first step of preparing a production task is to define its workflow (Figure 2). Usually a task consists of several stages each using different applications, databases, options, etc. The workflow describes the sequence of applications to be executed together with all the necessary application parameters. This includes software versions, application options, input and output data types. The workflow definitions are stored in the Production database to ensure data history information.

**2.5.2 Production jobs**

Once the workflow is defined, it can be used in a production run. The production run determines a set of data to be produced under same conditions. It instantiates a particular workflow together with some extra parameters like a number of events to be produced, specific application options for this run, a particular destination site for this run, etc.

The production run is split into jobs as units of the scheduling procedure. Each Production Agent request is served with a single job. The job is described in a XML file and contains all the necessary information for the Agent to successfully execute it.

All the production run parameters as well as job descriptions are stored in the Production database.

## 2.6 Production service

The Production service is an interface to the Production database allowing communicating with the Agents. It receives requests for jobs, checks the corresponding resource capabilities and then serves an outstanding job to the Agent.

The Production service has an XML-RPC server interface.

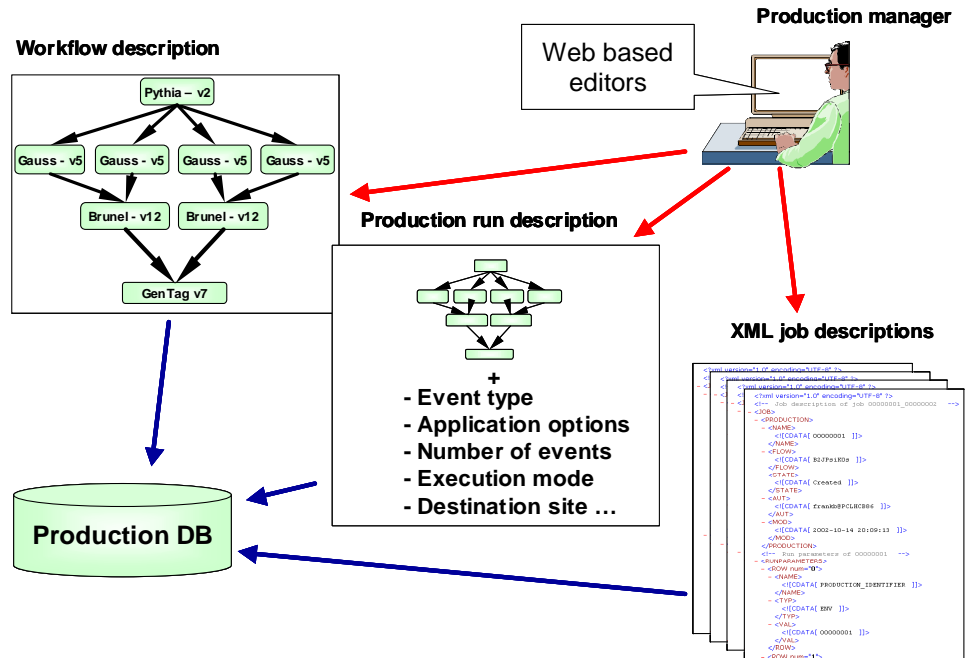

Figure 2: Definition of the production tasks

## 2.7 Monitoring service

Throughout the life cycle of a job its Agent sends messages to the Monitoring service to report the job progress. These messages are stored in the Production database and are visualized on a dedicated web page.

The Monitoring service has an XML-RPC server interface.

## 2.8 Bookkeeping service

When new datasets are produced they are registered by sending an XML dataset description to the Bookkeeping service. Here all the dataset descriptions are stored in a cache before they are checked by the production manager. After the check is done, the dataset metadata information is passed to the LHCb Bookkeeping database.

The Bookkeeping database is hosted by the CERN Oracle server. It can be interrogated by users using a dedicated web page with specialized forms. The output of the queries can be used directly in the user analysis jobs to specify input data.

The Bookkeeping service has an XML-RPC server interface.

## 2.9 Production Agent

The set of Agents running on all the production sites is the most important part of the system. It is their continuous work that allows to fill the computing resources with the LHCb production jobs thus realizing the "pull" job scheduling paradigm.

**TUAT006**

### 2.9.1 Agent setup on a production site

For a first time installation at a production site, a script is downloaded from a web page. This script creates the required directory structure and downloads the latest Production Agent software environment. Some local customizations are necessary. They are kept in a single script and concern mostly the interface to the local mass storage and to the local batch system. Subsequent releases of scripts are installed on top of currently installed ones, except the locally modified ones. No production application software is installed at this moment. This will be done later based on the requirements of a particular job. The installation procedure is very simple making the addition of new sites easy.

The Production Agent is invoked regularly at a production site to perform the following operations (Figure 3):

- When the occupancy of a local batch queue drops below a given level, the Production Agent interrogates the production service for new production requests.
- The Production Agent gets a job and checks if the required version of the software is available; if not it fetches it from the release area at CERN and installs it.
- The production agent submits a job to the local batch system.
- The job executes and writes its output to the local storage system.
- Upon completion of the job, the production agent ensures the output datasets are transferred to Castor at CERN and the bookkeeping database is updated.
- When the datasets have been successfully transferred, the replica information is added to the bookkeeping database.

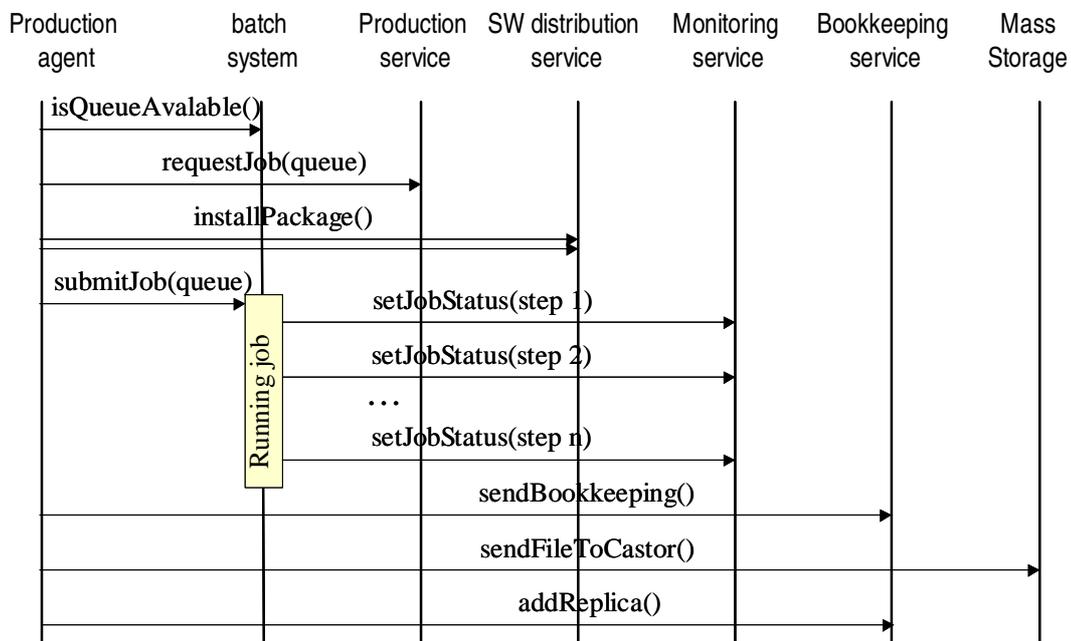

Figure 3: Production Agent operations

### 2.9.2 Agent implementation

The Agent is implemented as a set of classes written in Python. This allows to have a clean object-oriented design together with a rapid development environment. It uses XML-RPC protocol to communicate with central services. The entire Agent functionality is implemented using the facilities from the standard Python library.

The Agent runs as a daemon process or can be configured as a *cron* job. It requires an outbound IP connection for communicating with the central Production Service.

In the Agent functionality implementation special attention was paid to dealing with different types of failures. In particular, when the Agent fails to submit a job to the local batch system, the job is automatically rescheduled and can be picked up by another production site.

One of the most delicate operations in the job life cycle is the transfer of datasets since the volumes of data transmitted over the network are large. In case of transfer failures, the data stays cached on a production site and another attempt to their transfer is done during the next Agent invocation until the transfer is successful.

All the sensitive files, like job logs, metadata updates are also cached at the production site if even they are transferred to CERN storage to be made available through the bookkeeping web server.

**TUAT006**

## 3 USING DIRAC FOR THE LHCB DATA CHALLENGE

The first LHCb Physics Data Challenge took place in February-April 2003. The goal of this production run was to provide sufficient amount of data to evaluate the LHCb physics performance to be presented in the Technical Design Report (TDR). This was also the first time the DIRAC system was used in the full scale LHCb production.

The results of the Data Challenge can be summarized in the following numbers:
- 2 months of continuous running;
- 36600 jobs executed; each job was running 20 to 40 hours depending on the hosting CPU;
- 34000 jobs executed successfully, success rate is 92%; the failures were due to the LHCb software errors (2%) as well as due to problems on production sites (disk quotas, batch system) and while data transfers (6%);
- 250'000 datasets were produced;
- ~20 TB of data stored to the mass storage.

We have achieved the level of about 1000 jobs running simultaneously during several weeks in 18 LHCb production sites. The status of the production jobs was available at any moment through the monitoring web page as an interface to the production database. Summary plots were automatically generated to facilitate the production progress. The CPU time sharing among the production centers is shown in Figure 4.

During this production there were several software updates that were automatically installed by the system. This flexibility allowed to react promptly to the needs of the physics analysis and to study various LHCb detector configurations.

In the whole the DIRAC production system proved to be very stable and efficient in using all the computing resources available to the LHCb collaboration. It allowed achieving the goals of the Data Challenge ahead of schedule, that was conservatively based on our previous production experience.

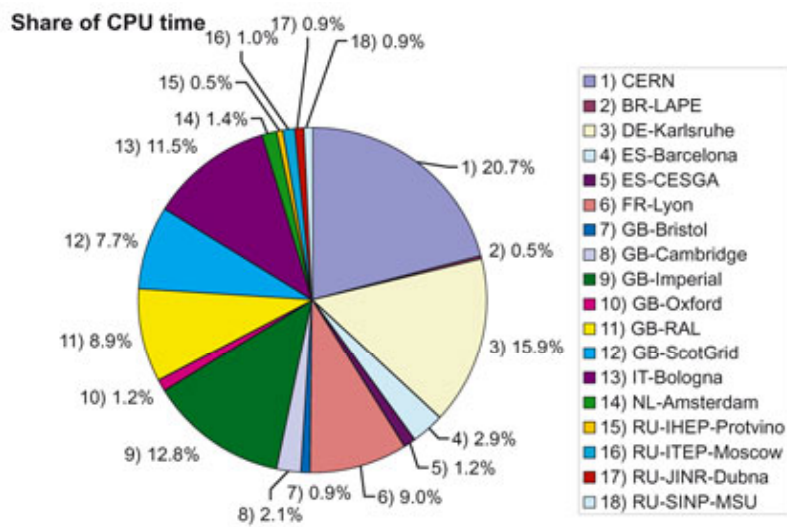

Figure 4: Sharing the CPU resources among different LHCb production sites

## 4 INTERFACING DIRAC TO THE DATAGRID TESTBED

As all the functionality required to steer the execution of jobs on a production site is encapsulated in a single component (the Production Agent) very little adaptation was necessary to integrate the DataGRID into DIRAC. As far as DIRAC is concerned, the DataGrid is just yet another remote center.

### 4.1 Deploying Agents on the DataGRID

A standard Production Agent running on the DataGRID portal, i.e. the host with an installed Grid user interface, ensures the job submission to the DataGRID (Figure 5). The same LHCb production service is used to get the jobs from. Then the job XML description together with an Agent installation script is packed into a InputSandbox detailed by the job JDL file. The so specified job is submitted the DataGRID Resource Broker. Upon starting the job on a Worker Node (WN), the Production Agent software is first installed thus effectively turning the WN into a LHCb production site. From this point onwards all the operations necessary to steer the job execution are the same as is described above for any other LHCb site.

The Agent running on the WN first checks and installs the applications software. The software is installed either on the closest Storage Element (SE) or in the job current directory. The latter is done if files in the closest SE are not available via a POSIX *open* call.

The Agent sends messages to the monitoring service to trace the job execution progress.

**TUAT006**

In the end of the job execution the produced data sets are copied to the Castor mass storage system at CERN. For that we use the Replica Manager software. In this case a dataset is copied to a SE situated at CERN configured in such a way that each dataset copy triggers execution of a GDMP script that saves the newly arrived file to the Castor storage. Doing so, we achieved the goal of easy access to data produced in the DataGRID by users outside the DataGRID. The bookkeeping information about the datasets produced is provided in both DataGRID Replica Catalog and in the LHCb bookkeeping database.

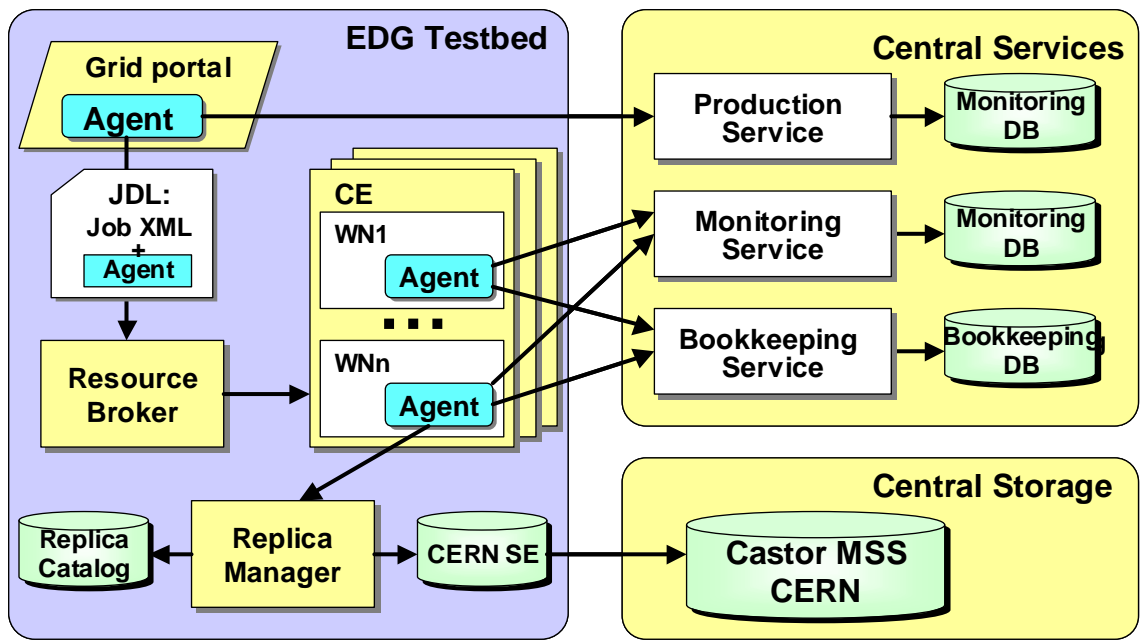

Figure 5: Running DIRAC jobs on the DataGRID

## 4.2 Results of the tests

A set of about 2000 test jobs were executed in the DIRAC/DataGRID setting in February-March 2003. The jobs differed by their time length ranging from 20 minutes to 50 hours on a 1GHz CPU. We have encountered many problems most of which are coming from the instabilities of the middleware available at this moment. The job success rate was between 25% and 60% depending on the job durability (the longer the jobs the higher their probability to fail). Nevertheless, we have demonstrated that the DataGRID can be included into the DIRAC production procedure. Standard LHCb production jobs were used in the tests without simplification to accommodate possible DataGRID shortcomings. The data produced are available for the LHCb users outside the Grid and can be traced both through the LHCb bookkeeping database and with the DataGRID replica catalogue.

## 5 DISCUSSION

The use of DIRAC in the LHCb MC production as well as tests with the DataGRID testbed revealed a number of problems that we will have to resolve in the future and/or can be addressed by the Grid middleware developers. Here we mention some of them and we do not address all the well-known problems of the existing middleware.

## 5.1 Application software installation

During the production it is often the case that the application software is getting updated because of the bugs found or a new crucial bit of code is added. The new versions of the software should be immediately available on all the production sites. In DIRAC this is achieved by automatic installation in the local disk area that is write-accessible for the Production Agent. In case of running the DIRAC jobs on the DataGRID the standard place where VO software is installed is usually not accessible for writing. Therefore, the closest SE is used for the installation which is not a satisfactory solution. In some cases the SE is not reachable via POSIX calls and then all the application libraries should be copied to the working directory before the execution starts. This is a clear waste of resources. Therefore, a standard middleware API for the VO software installation is very desirable. This should include methods for software installation and uninstallation, registration of the installed packages, getting information on the available software, tracking dependencies between the packages.

**TUAT006**

## 5.2 WN IP connectivity

The experience of the DIRAC operation showed a great importance of the possibility for the running jobs to send messages to the central services. This is the only way to have fine grain information of the job progress that allows reacting quickly in case of problems. It requires however the outbound IP connectivity of the working nodes that is highly debatable because of the security concerns. Having a dedicated Agent running on the production site gatekeeper host can solve this problem by channeling the messages through the Agent. The gatekeeper host has normally access to the outside WAN and can be also reached from outside network. Therefore, this allows sending messages to the running job and opens the possibility to run interactive tasks. This is very desirable for the future data analysis models. In particular, the possibility of running experiment-dedicated daemons on a computing element can be considered for inclusion in the eventual grid middleware.

## 6 CONCLUSIONS

The DIRAC system is routinely used now in the LHCb Collaboration for massive MC data productions. It allowed to efficiently use all the available computing resources with a minimum intervention of the local production site managers. The Physic Data Challenge 1 that took place in February-April 2003 was a success justifying the DIRAC conceptual choices.

The modular client/server design with central services and a set of distributed Agents allowed us to rapidly develop different components once the communication protocols were fixed. This approach of having various independent grid services orchestrated by agents in charge of scheduling the jobs is very promising. Therefore, we are looking forward to the introduction of the Open Grid Services Architecture (OGSA) as proposed recently by the Global Grid Forum [5].

The "pull" job scheduling strategy proved to be adequate for the MC production tasks. We are looking forward now to extend the DIRAC functionality to the analysis type tasks, i.e. the tasks that need a selection of input data and which are not planned regularly. The AliEn distributed computing system [6] uses a similar approach. This system has been in production for more than a year now. It was used with success by multiple experiments both for the MC production and analysis tasks.

DIRAC system does not rely on Grid technology; with DataGrid being used as a separate production center. DIRAC thus provides LHCb with the flexibility to use different types of Grids and non-Grid farms simultaneously.

## Acknowledgments

The authors would like to thank the LHCb site administrators for their kind assistance in the deployment of the production software and quick resolution of the operational problems.

We wish also to acknowledge a very good cooperation of the DataGRID middleware developers and especially that of the Integration Team.

## References

[1] The LHCb Collaboration, LHCb: Technical Proposal, CERN-LHCC-98-004;

[2] The DataGRID project, http://eu-datagrid.web.cern.ch/eu-datagrid;

[3] G. Barrand, et al, GAUDI - A Software architecture and framework for building hep data processing applications, Comput. Phys. Comm, 140(2001)45;

[4] CMT code management tool, http://www.cmtsite.org;

[5] Global Grid Forum, http://www.ggf.org;

[6] The AliEn project, http://alien.cern.ch;